\documentclass[prb,showpacs,preprintnumbers,preprint,superscriptaddress]{revtex4}

\usepackage{graphicx}
\usepackage{dcolumn}
\usepackage{bm,hyperref}
\usepackage{amsmath}
\usepackage{amsfonts}
\usepackage{amssymb}%

\begin{document}
\title{May optical fishnet metamaterials be described by effective material parameters?}
\author{Christoph Menzel}
\affiliation{Institute of Solid State Theory and Optics,
Friedrich-Schiller-Universit\"at Jena, Germany}
\author{Thomas Paul}
\affiliation{Institute of Solid State Theory and Optics,
Friedrich-Schiller-Universit\"at Jena, Germany}
\author{Carsten Rockstuhl}
\affiliation{Institute of Solid State Theory and Optics,
Friedrich-Schiller-Universit\"at Jena, Germany}
\author{Thomas Pertsch}
\affiliation{Institute of Applied Physics,
Friedrich-Schiller-Universit\"at Jena, Germany}
\author{Sergei Tretyakov}
\affiliation{Department of Radio Science and Engineering / SMARAD,
Helsinki University of Technology, P.O. Box 3000, FI-02015 TKK,
Finland}
\author{Falk Lederer}
\affiliation{Institute of Solid State Theory and Optics,
Friedrich-Schiller-Universit\"at Jena, Germany}

\begin{abstract}
Although optical metamaterials that show artificial magnetism are
mesoscopic systems, they are frequently described in terms of
effective material parameters. But due to intrinsic nonlocal (or
spatially dispersive) effects it may be anticipated that this
approach is usually only a crude approximation and is physically
meaningless. In order to study the limitations regarding the
assignment of effective material parameters, we introduce a technique
to retrieve the frequency-dependent elements of the effective
permittivity and permeability tensors for arbitrary angles of
incidence and apply the method exemplarily to the fishnet metamaterial.
It turns out that for the fishnet metamaterial, genuine effective
material parameters can only be introduced if quite stringent constraints
are imposed on the wavelength/unit cell size ratio. Unfortunately
they are only  met far away from the resonances that induce a magnetic
response  required for many envisioned applications of such a fishnet
metamaterial. Our work clearly indicates that the mesoscopic nature
and the related spatial dispersion of contemporary optical
metamaterials that show artificial magnetism prohibits the
meaningful introduction of conventional effective material
parameters.

\end{abstract}

 \maketitle

The introduction of nanostructured metamaterials (MMs) into optics
potentially opens the door to a fairly comprehensive control of
light propagation. During the past several years much effort has
been devoted to achieve this goal and two major research fields may
be distinguished. On the one hand, advances in nanotechnology
provide ever smaller and more complex structures which constitute
quite involved nanostructured media. On the other hand, optics in
media with unprecedented effective material equations has been
investigated purely theoretically and novel effects have been
revealed \cite{ComplexMediaBook}. The desirable assignment of
effective material parameters to a specific MM would bridge the gap
between both approaches and allow to link a fabricated structure to
a particular effective constitutive relation
\cite{ParameterRetrievalNormal, PRBianisotrop, PRChiral}.
Furthermore it would appreciably facilitate the description of light
propagation in optical MMs and their combination with other optical
materials because canonical Maxwell boundary conditions could be
applied. If we wish to understand under a MM an artificial medium
made of periodically or non-periodically arranged metaatoms which
allows to control the properties of light propagation predominantly
by the chosen geometry of the metaatoms, many different MMs can be
envisioned which all have peculiar aspects if effective properties
shall be designed \cite{Narimanov,Schweizer,NatMatValentine}.
To avoid any misunderstanding, we wish to restrict
our considerations in the following on metamaterials that show the
effect of an artificial magnetism and which shall operate at optical
frequencies. Such property is often at the focus of interest since
the media would enable optical phenomena that contradict our common
perception of how light propagates.

For such special type of MM, there is, however, a serious issue which might prevent this
simplified description in terms of effective material parameters. In general, typical optical MMs are
mesoscopic where the vacuum wavelength is only a few times larger
than the unit cell. In such systems the optical response may be
reasonably described by induced currents, which nonlocally depend on
the electric field. In Fourier space this leads to a spatially
dispersive conductivity, as discussed in the work of Serdyukov
\textit{et al.} [\onlinecite{TretyakovBook}]. At this stage it is
not required to distinguish between polarization
$(\sim\frac{\partial}{\partial{t}}\mathbf{P})$ and magnetization
currents $(\sim\nabla\times\mathbf{M})$, however, this becomes
important if either of them becomes resonant in the nanostructure.
Provided that this spatial dispersion is weak, the constitutive
relation between $\mathbf{j}$ and $\mathbf{E}$ can be expanded up to
the second order. Since the fields $\mathbf{D}$ and $\mathbf{H}$
cannot be defined uniquely, spatial derivatives of $\mathbf{E}$ may
be replaced in favor of $\mathbf{B}$. As a result, two constitutive
relations $\mathbf{D}(\mathbf{E},\mathbf{B})$ and
$\mathbf{H}(\mathbf{E},\mathbf{B})$ emerge with tensorial, but only
frequency dependent coefficients \cite{TretyakovBook}. First order
terms lead to magnetoelectric ($\mathbf{E}
\rightleftarrows\mathbf{H}$) coupling (bianisotropy, chirality).
Second order terms to anisotropic, but spatially nondispersive
relations between both $\mathbf{D}$ and $\mathbf{E}$ and
$\mathbf{H}$ and $\mathbf{B}$. To sum up, these so-called
bianisotropic constitutive relations are the most general ones for a
weak spatially dispersive conductivity in a nanostructured material.
From a physical point of view it is appealing that the nonlocal
relation between the electric field and the induced currents is the
very source for the effective chiral and magnetic
($\hat{\mu}(\omega)$) properties of MMs. From a technical point of
view, with these spatially nondispersive constitutive relations at
hand, standard boundary conditions \cite{Jackson} can be used to
solve macroscopic Maxwell's equations in layered media. This has a
big advantage compared to the rather involved procedure for
spatially dispersive constitutive relations which require the use of
so-called additionally boundary conditions \cite{Agranovich}. In
mirror-symmetric (nonchiral) media first order terms in the
expansion vanish and the magneto-electric coupling disappears. The
MM may then be described by two material tensors
$\hat{\varepsilon}(\omega)$ and $\hat{\mu}(\omega)$.\\
Here we aim at introducing a simple criterion that tells us if this condition is fulfilled.
In general, one has to develop an approach to retrieve these tensor elements
from reflection/transmission data. However, it will turn out that the very
calculation of these parameters is not required. Only if the criterion is
fulfilled one has to proceed with the retrieval algorithm to calculate
effective parameters which are then independent of the incidence angle and may
be termed effective \emph{material parameters}. On the other hand, if the
criterion is not fulfilled the assignment of an effective permittivity
$\hat{\varepsilon}(\omega)$ and permeability $\hat{\mu}(\omega)$ is pointless.
This means physically that the assumption of weak spatial dispersion is
violated and the effective material parameters would become spatially
dispersive making the approach used inconsistent. Thus the aim of this work is
not the retrieval of parameters but to evaluate if a certain MM may be
described by effective material parameters.
\\
But in any case the recently introduced retrieval approach for isotropic
materials at arbitrary incidence \cite{PRIsotrop} has to be generalized towards
anisotropic media. The main advantage of the present approach is that all
tensor elements can be determined without requiring explicitly that the
propagation direction coincides with a crystallographic axis. Hence, the
approach may be applied to all
currently fabricated MMs.\\
To start with, we assume that the metallic structures (split ring,
fishnet, etc.) involved are reciprocal and not intrinsically
magnetic ($\mu_{\textrm{metal}}=1$). Their response to the
electromagnetic field can be completely described by a current
nonlocally induced by the electric field as \cite{TretyakovBook}
\begin{equation}
\mathbf{j}(\mathbf{r},\omega)=\int\limits_{V}\widehat{{R}}{(\mathbf{r},\mathbf{r^{\prime}},\omega)\mathbf{E}(\mathbf{r^{\prime}},\omega)d\mathbf{r^{\prime}}},\label{j}
\end{equation}
where the dyadic
$\widehat{{R}}{(\mathbf{r},\mathbf{r^{\prime}},\omega)}$ describes
the nonlocal response of the medium. However this approach is not
practical. For weak nonlocality (or spatial dispersion) one may
rather expand Eq.(\ref{j}) up to the second order. For improving the
readability of the paper we provide our derivation in following the
lines in Ref. \onlinecite{TretyakovBook}. We obtain
\begin{equation}
j_{k}(\mathbf{r},\omega)\approx i\omega\left[  a_{kl}E_{l}+b_{klm}%
\frac{\partial E_{l}}{\partial x_{m}}+c_{klmn}\frac{\partial
E_{l}}{\partial
x_{m}\partial x_{n}}+...\right]  , \label{j_2}%
\end{equation}
where Einstein notation is applied and $b_{klm}=-$ $b_{lkm}$. The
factor $i\omega$ has been introduced for convenience, so that the
terms in the brackets represent the induced polarization. Now the
constitutive relations read as
\begin{align}
D_{k}(\mathbf{r},\omega)  &  =\left(  \varepsilon_{0}\delta_{kl}%
+a_{kl}\right)  E_{l}+b_{klm}\frac{\partial E_{l}}{\partial x_{m}}%
+c_{klmn}\frac{\partial^2 E_{l}}{\partial x_{m}\partial x_{n}},\label{cr_1}\\
\mathbf{H}\mathbf{=B/}  &  \mu_{0}. \label{cr_2}%
\end{align}
The first term leads to the anisotropic permittivity
$\varepsilon_{kl}=\delta_{kl}+a_{kl}/\varepsilon_0$. The second term accounts
for magnetoelectric coupling and vanishes for media with three orthogonal
planes of mirror symmetry, i.e. media that are purely anisotropic in the
quasi-static limit. To proceed we require the coefficients $c_{klmn}$ to obey
the following relation:
\begin{equation}
c_{klmn}\frac{\partial^2 E_{l}}{\partial x_{m}\partial x_{n}}\stackrel{!}{=}[\nabla\times(\hat{\gamma}\nabla\times\mathbf{E})]_k\label{EQ_GAMMA}%
\end{equation}
where the components $\gamma_{ij}$ are related to the coefficients
$c_{klmn}$. Since $c_{klmn}=c_{lkmn}$ holds, the tensor
$\hat{\gamma}$ is symmetric ($\gamma_{ij}=\gamma_{ji}$).\\
Because Maxwell's equations are invariant with respect to the transformations
[\onlinecite{TretyakovBook}] $
\mathbf{D}^{\prime}=\mathbf{D}+\nabla\times\mathbf{Q},
{H}^{\prime}=\mathbf{H}-i\omega\mathbf{Q} $ we can rewrite Eqs. \ref{cr_1} and
\ref{cr_2} in using
$\mathbf{Q}(\mathbf{r},\omega)=-\hat{\gamma}\nabla\times\mathbf{E}=i\omega\hat{\gamma}\mathbf{B}$
to obtain the ultimate constitutive relations
\begin{equation}
\mathbf{D}(\mathbf{r},\omega)=\varepsilon_{0}\hat{\varepsilon}(\omega
)\mathbf{E}(\mathbf{r},\omega),\hspace{3mm}  \mathbf{B}(\mathbf{r},\omega)=\mu_{0}\hat{\mu}(\omega)\mathbf{H}%
(\mathbf{r},\omega).\label{EQ_D}%
\end{equation}
These equations represent our point of departure. They reflect that
a mesoscopic metallic structure with a weak nonlocal response (weak
spatial dispersion) can be likewise treated as an effective
homogeneous, anisotropic but magnetic medium where the magnetic
properties merely originate from the nonlocal response
$\widehat{{R}}(\mathbf{r},\mathbf{r^{\prime}},\omega)$:
\begin{equation} \mu_{kl}=(\delta_{kl}-\mu_0\omega^2\gamma_{kl})^{-1}.
\end{equation}
Only if the constitutive relations of a medium without bianisotropy can be cast
in the form of Eqs.~(\ref{EQ_D}), the usual boundary conditions are applicable
where the tangential components $E_{t}$ and $H_{t}$ as well as the normal
components
$D_{n}$ and $B_{n}$ are continuous.\\
The strategy of our work is as follows: We develop a retrieval
algorithm for anisotropic, homogeneous and local media which relies
on the rigorously calculated reflection/transmission data from a MM
slab for transverse wave vectors that extend even into the
evanescent domain. Within this algorithm we identify a quantity
$\alpha(\omega)$ which contains the essential information. If this
quantity depends only on frequency and not on the wave vector, the
effective permittivity and permeability tensor elements, not
calculated at this step, will exhibit the same feature and can be
calculated in a further step and assigned to the metamaterial. If
this is not the case the ratio unit cell size/wavelength has
to be decreased until this criterion is met. Here we restrict
ourselves to media where both material tensors can be simultaneously
diagonalized as
\begin{equation}
\hat{\varepsilon}=\textrm{diag}\{\varepsilon_x,\varepsilon_y,\varepsilon_z\}
,\hspace{2mm}\hat{\mu}=\textrm{diag}\{\mu_x,\mu_y,\mu_z\}. \label{EQ_Tensors}%
\end{equation}
\begin{figure}[h]
\centering
\includegraphics[width=70mm,angle=0] {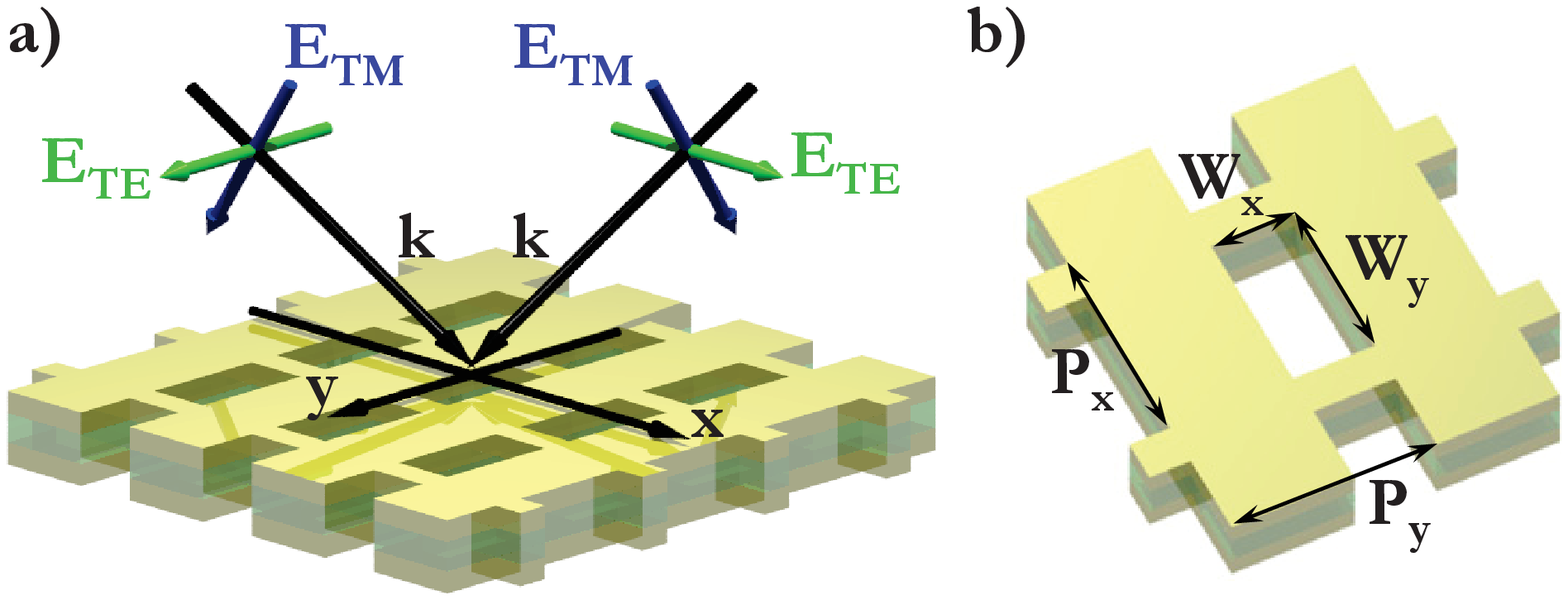}
\caption[Submanifold] {(color online) a) Schematic view of the
single fishnet layer together with the four principal directions for
the retrieval. b) Unit cell of the fishnet with
$P_x=P_y=600\,\textrm{nm}$, $W_x=284\,\textrm{nm}$,
$W_y=500\,\textrm{nm}$ embedded in air. The thicknesses of the
silver and the intermediate $\textrm{MgF}_2$ ($n=1.38$) layer are
$d_{\textrm{Ag}}=45\,\textrm{nm}$ and $d=30\,\textrm{nm}$,
respectively.} \label{FIG_KOS}
\end{figure}
We align the coordinate system and therefore all interfaces to the
crystallographic axis of the effective anisotropic medium. The
incident light shall consist of monochromatic plane waves whose
wavevector is perpendicular to at least one coordinate axis. Then
the eigenmodes of the medium can be decomposed into decoupled TE and
TM modes and the reflection/transmission problem is equivalent to
that of an isotropic medium. The only difference is the different
propagation constant for each eigenmode. For a particular
eigenpolarization the reflection and transmission coefficients in
terms of the normal wavevector component have the same form as in
the isotropic case. For varying the incidence plane and the
polarization only certain quantities have to be exchanged as
indicated in Tab.~\ref{TABLE_EXCHANGE}.
 The transmission and
reflection coefficients for the electric field read as
\begin{equation}
T(k,\xi)=\frac{2k_{s}\xi A}{\xi(k_{s}+k_{c})\cos(kd)-i(\xi^{2}+k_{s}k_{c}%
)\sin(kd)} \label{TransCoeffSimpler}%
\end{equation}%
\begin{equation}
R(k,\xi)=\frac{\xi(k_{s}-k_{c})\cos(kd)+i(\xi^{2}-k_{s}k_{c})\sin(kd)}%
{\xi(k_{s}+k_{c})\cos(kd)-i(\xi^{2}+k_{s}k_{c})\sin(kd)}
\label{ReflCoeffSimpler}%
\end{equation}
where the following abbreviations have been used:
\begin{align}
k_{s,c}=\alpha^{s,c}k_{z}^{s,c},~~k=k_{z}^{f},~~A_{TE}=1,~~A_{TM}=\sqrt{\frac{\varepsilon^{s}\mu^{c}}{\varepsilon^{c}\mu^{s}}}
\label{EQ_IntroXi}
\end{align}
\begin{equation}
\text{with}\,~~\xi=\alpha^{f}k_{z}^{f},~~\text{where}\,~~k_{z}^{i}=\sqrt{\frac{\omega^{2}}{c^{2}}\beta^{i}-k_{t}^{2}\cdot\gamma^{i}}
\label{DispRelAnisotropic}%
\end{equation}
is the normal component of the wavevector in medium
$^{\prime}i^{\prime}$ and $k_{t}=(k_{x,}k_{y})$ is its conserved
tangential component.
The superscripts $i\in\{s,f,c\}$ denote substrate, film and
cladding. In what follows we assume that substrate and cladding are
isotropic and nonmagnetic where $\alpha_{\mathrm{TE}}^{s,c}=1$,
$\alpha_{\mathrm{TM}}^{s,c}=1/\varepsilon^{s,c},\beta^{s,c}\equiv
\varepsilon^{s,c}$ and $\gamma^{s,c}\equiv1$. For the sake of
clarity we drop the superscript $^{\prime}f^{\prime}$ and write
$\alpha^{f}(\omega)=\alpha(\omega)$,$~\beta^{f}(\omega)$ $=\beta
(\omega)$ and $\gamma^{f}(\omega)=\gamma(\omega)$. These
coefficients are related to different combinations of tensor
components of the permittivity and permeability depending on the
polarization and the incidence plane, see
Table~\ref{TABLE_EXCHANGE}.
\begin{table}[h]
\extrarowheight3pt
\begin{tabular}
[c]{c||c|c||c|c}%
$~~$ & \multicolumn{2}{c||}{\textbf{TE}} & \multicolumn{2}{c}{\textbf{TM}%
}\\\hline & $~k_{x}=0~$ & $~k_{y}=0~$ & $~k_{x}=0~$ & $~k_{y}=0~$\\\hline
$\alpha$ & $1/\mu_{y}$ & $1/\mu_{x}$ & $1/\varepsilon_{y}$ & $1/\varepsilon
_{x}$\\\hline $\beta$ & $\varepsilon_{x}\mu_{y}$ & $\varepsilon_{y}\mu_{x}$ &
$\varepsilon _{y}\mu_{x}$ & $\varepsilon_{x}\mu_{y}$\\\hline
$\gamma$ & $\mu_{y}/\mu_{z}$ & $\mu_{x}/\mu_{z}$ & $\varepsilon_{y}%
/\varepsilon_{z}$ & $\varepsilon_{x}/\varepsilon_{z}$
\end{tabular}
\caption{Substitution table for the relevant coefficients depending on the
polarization and the incidence plane.}%
\label{TABLE_EXCHANGE}%
\end{table}
For the sake of brevity the frequency dependence will be kept in mind but not
explicitly written in the following.

\begin{figure}[h]
\centering
\includegraphics[width=84mm,angle=0] {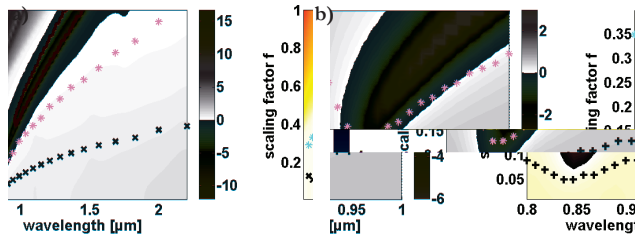}
\caption[Submanifold]{(color online) a) Real part of the propagation constant
$k$ (in $\mu m^{-1}$) of the fishnet structure for normal incidence. Additionally the lines of
constant relative deviation of $\alpha$ are shown (black cross - $0.1\% $, cyan
asterisk - $5\%$). The relative deviation is calculated for varying $k_t=k_y$ in TE polarization
($\alpha=\mu_y^{-1}$) being the preferential operating polarization state for
the fishnet. The green solid line indicates the region of $\Re(\mu)<0$. b) zoomed domain of interest from a).}
\label{FIG_RealKErrorAlpha}%
\end{figure}

The effective material slab is then fully characterized by the parameters $k$ and $\xi$. Note that throughout the paper $k$ is the normal
component of the wavevector in the slab. By inverting
Eqs.~(\ref{TransCoeffSimpler}) and (\ref{ReflCoeffSimpler}) one obtains
\begin{equation}
kd=\pm\arccos\left(  \frac{k_{s}(1-R^{2})+k_{c}(T/A)^{2}}{(T/A)[k_{s}%
(1-R)+k_{c}(1+R)]}\right)  +2m\pi\label{Retrieved_KF}%
\end{equation}
with $m\in\mathbb{Z}$ and
\begin{equation}
\xi=\pm\sqrt{\frac{k_{s}^{2}(R-1)^{2}-k_{c}^{2}(T/A)^{2}}{(R+1)^{2}-(T/A)^{2}%
}}. \label{Retrieved_ZF}%
\end{equation}
The sign of $k$ and $\xi$ and the branch order $m$ are determined by
the usual physical constraints \cite{PRIsotrop}. The quantities $k$
and $\xi$ can be uniquely determined and are the final
\emph{effective wave parameters}. They are independent of the
thickness $d$ of the slab, provided that they already converged
towards the bulk data \cite{ImagingPropThomas}. Then these wave parameters
must coincide with those provided by the dispersion relation of the
fundamental Bloch mode \cite{LightPropFishnet}. It is evident that
these effective wave parameters describe properties of the
fundamental Bloch mode in the infinite lattice formed by
periodically arranging a single fishnet layer \cite{prb07,meta07}.
They still depend on the propagation direction, the angle of
incidence and the polarization state as in any anisotropic medium.
\\
Now the criterion central to this work can be formulated. Since
$\alpha(\omega)=\xi/k$ (see Eq.~\ref{DispRelAnisotropic}) is related to the
effective material parameter tensor components (see
Tab.~\ref{TABLE_EXCHANGE}) it has to be independent of the angle of
incidence although $\xi$ and $k$ will strongly depend on it.
Actually this is a very simple criterion that we can use to evaluate
the validity of the effective material approach just by calculating
$\xi$ and $k$ from the transmission/reflection data by using
Eqs.~(\ref{Retrieved_KF}) and (\ref{Retrieved_ZF}).
\\
Only in the case that the above criterion is fulfilled the very
parameter retrieval can be performed, otherwise their assignment is
pointless. This parameter retrieval may then be performed as
follows. If the wave parameters $k$ and $\xi$ are determined for a
certain polarization and incidence plane, the parameter $\alpha$ is
given by $\alpha=\xi/k$. The parameter $\beta$ follows then from
Eq.~(\ref{DispRelAnisotropic}) at normal incidence which reads as
$k=\frac{\omega}{c}\sqrt{\beta}$. With $\beta$ being constant, the
remaining parameter $\gamma$ can be retrieved as a function of
$k_{t}$ from Eq.~(\ref{DispRelAnisotropic}) as
$\gamma=\frac{k^{2}-\frac{\omega^{2}}{c^{2}}\beta}{k_{t}^{2}}.$ Now
having $\alpha$, $\beta$,and $\gamma$, the classification in
Tab.~\ref{TABLE_EXCHANGE} can be used to retrieve the effective
material parameters for the respective polarization and incidence
plane. It can be recognized that each parameter can be retrieved
twice; providing a possibility to double-check results. If both
results agree the model of weak spatial dispersion holds and the effective material parameters are meaningful.\\\
The outlined procedure is exemplarily applied to the fishnet MM.
Geometrical parameters are given in Fig.~\ref{FIG_KOS}. The metal
layers are assumed to be made of silver \footnote{
$\varepsilon=1-{\omega_p^2}/{\omega(i\gamma+\omega)}$ with
$\omega_p=1.37\cdot10^{16}\,\textrm{s}^{-1}$ and
$\gamma=8.5\cdot10^{13}\,\textrm{s}^{-1}$}. To identify the limits of
the effective parameter description, we varied the wavelength to
unit cell size ratio by scaling the structure and calculated
$\alpha=\xi/k$. Hence, all parameters defined in Fig.~\ref{FIG_KOS}
are scaled by a factor $f$. Results of the numerical simulations are
summarized in Fig.~\ref{FIG_RealKErrorAlpha} where the real part of
the propagation constant $k$ is shown as a function of the
wavelength and the scaling factor for normal incidence. The dark-colored
area exhibits a negative propagation constant and can thus be
considered magnetically active (antisymmetric plasmon resonance). In
the light-colored area the fishnet exhibits a plasma-like effective
permittivity with negligible dispersion in the permeability.
The green solid line indicates the region of $\Re(\mu)<0$ at normal incidence, i.e. the region of double negativity.
It is evident that the smaller the structure the smaller the resonance
wavelength and the weaker the resonance strength. For very small scaling
factors ($f < 0.1$, i.e. unit cell size of less than 60 nm) where the quasi-static limit is reached, the
resonance is almost wavelength independent but also tends to disappear; being in
agreement with findings for split-rings
\cite{Ishikawa,Zhou}.
\\
Having identified the area where resonances are occurring it is now
interesting to disclose where the effective parameter description
($\alpha$ must be invariant) may be applied. To this end we
calculated for discrete wavelengths the parameter $\alpha$ depending
on the transverse wavevector ($k_t=0..1.2k_0$) including at least
partially the evanescent spectrum. The relative deviation
$(\max|\alpha(k_t)-\alpha(0)|)/|\alpha(0)|)$ serves as a measure to
characterize the variation of $\alpha$ and is displayed in
Fig.\ref{FIG_RealKErrorAlpha}. Two bounds for this deviation are
considered,
 $0.1\%$ (almost ideal assignment of effective material parameters possible)
 and $5\%$ (assignment of effective parameters might be still feasible).
 Clearly, close to the resonance this deviation
is strongest for a fixed scaling factor. It is evident that for a
required deviation of $0.1\%$ effective material parameters can be
only introduced when there are no magnetic resonances (no
lefthandness). Since an effective description holds only for almost
invariant $\alpha$ this deviation should be as small as possible but
values of $5\%$ might be tolerable at most. This condition requires
a scaling factor of about $0.15$ in the resonance region resulting
in a wavelength to cell size ratio of $\lambda/P > 10$. Evidently,
the resonance strength is very weak in this domain leading to a
non-magnetic response of the material (lefthandness occurs only
because of the large imaginary parts). This is consistent with the
assumption that such small structures can be described in the
quasi-static limit where no magnetic response is observed. Hence,
the result of our studies is quite discouraging, namely: a
sufficiently strong magnetic response ($\Re(\mu)<0$) requires a certain minimum
unit cell size/wavelength ratio (about 1:4 in case of the fishnet),
but this mesoscopic structure must not be described by conventional
frequency-dependent effective permittivity and permeability tensors.
We have proven this for a fishnet structure, but since all present
optical MMs rely on similar resonances we conclude that this
tendency may hold in general.
\\
To sum up, based on the assumption of a weakly spatially dispersive
conductivity in MM unit cells we have developed a method to
verify/falsify an effective anisotropic medium description of MMs.
We have shown that a prototypical magnetically active, and thus
potentially negative index, material, namely the fishnet, cannot be
described as a homogeneous anisotropic medium in the relevant resonance
region. By varying the ratio of wavelength to cell size we have
elaborated the limitations of the weak spatial dispersion
assumption. There is a trade off: If the spatial dispersion is weak
and the material parameters have the usual meaning, the
antisymmetric plasmonic resonance, which is responsible for magnetic
activity, is also weak or disappears. Our work clearly indicates
that for optical MMs the commonly assigned effective parameters do
not have the physical meaning of conventional material parameters.\\
This work was partially supported by the German Federal Ministry of
Education and Research (Metamat), by the ProExzellence Initiative of
the State of Thuringia, and FP7 CSA ECONAM and the NANOGOLD project.

\end{document}